\documentclass{article}

\usepackage{PRIMEarxiv}

\usepackage[utf8]{inputenc} 
\usepackage[T1]{fontenc}    
\usepackage{hyperref}       
\usepackage{url}            
\usepackage{booktabs}       
\usepackage{amsfonts}       
\usepackage{nicefrac}       
\usepackage{microtype}      
\usepackage{lipsum}
\usepackage{fancyhdr}       
\usepackage{graphicx}       
\graphicspath{{media/}}     
\usepackage[authoryear]{natbib}
\usepackage{float}

\providecommand{\tightlist}{%
  \setlength{\itemsep}{0pt}\setlength{\parskip}{0pt}}

\title{Memshare: Memory Sharing for Multicore Computation in R with an Application to Feature Selection by Mutual Information using PDE}

\author{
  Michael C. Thrun \\
  University of Marburg, \\
  Mathematics and Computer Science, D-35032 Marburg\\
  IAP-GmbH Intelligent Analytics Projects \\
  In den Birken 10A, 29352 Adelheidsdorf\\
  \url{https://www.iap-gmbh.de}\\
  \textit{ORCiD: \href{https://orcid.org/0000-0001-9542-5543}{0000-0001-9542-5543}}\\
  \href{mailto:mthrun@informatik.uni-marburg.de}{\nolinkurl{mthrun@informatik.uni-marburg.de}}
   \And
  Julian Märte \\
  University of Marburg, \\
  Mathematics and Computer Science, D-35032 Marburg\\
  IAP-GmbH Intelligent Analytics Projects \\
  In den Birken 10A, 29352 Adelheidsdorf\\
  \textit{ORCiD: \href{https://orcid.org/0000-0001-5451-1023}{0000-0001-5451-1023}}\\
  \href{mailto:j.maerte@iap-gmbh.de}{\nolinkurl{j.maerte@iap-gmbh.de}}
}

\begin{document}
\maketitle

\begin{abstract}
We present memshare\footnote{The Software package is published as a CRAN package under \cite{thrunmaerteR}}, a package that enables shared memory multicore computation in R by allocating buffers in C++ shared memory and exposing them to R through ALTREP views. We compare memshare to SharedObject (Bioconductor) discuss semantics and safety, and report a 2× speedup over SharedObject with no additional resident memory in a column wise apply benchmark. Finally, we illustrate a downstream analytics use case: feature selection by mutual information in which densities are estimated per feature via Pareto Density Estimation (PDE). The analytical use-case is an RNA seq dataset consisting of N=10,446 cases and d=19,637 gene expressions requiring roughly n\_threads * 10GB of memory in the case of using parallel R sessions. Such and larger use-cases are common in big data analytics and make R feel limiting sometimes which is mitigated by the addition of the library presented in this work.
\end{abstract}


\section{Introduction}\label{introduction}

Parallel computing in R is usually realized through PSOCK or FORK clusters, where multiple R processes work in parallel \citep{Rparallel2025}, \citep{doparallel2025}. A practical issue arises immediately: each worker process receives its own private copy of the data. If a matrix consumes several gigabytes in memory, then spawning ten workers results in ten redundant copies, potentially exhausting system memory and severely reducing performance due to paging. This overhead becomes especially prohibitive in genomics or imaging, where matrices of tens of gigabytes are commonplace. Copying also incurs serialization and deserialization costs when transmitting objects to workers, delaying the onset of actual computation.

Shared memory frameworks address this issue by allowing workers to view the same physical memory without duplication. Instead of copying the whole object, only small handles or identifiers are communicated, while the underlying data is stored once in RAM. This enables efficient multicore computation on large datasets that would otherwise be infeasible.

ALTREP (short for ALTernative REPresentations) is a framework in R that allows vectors or matrices to present an alternative backend for their storage while still behaving like ordinary R objects. For example, ALTREP can be used to represent a sequence vector without actually allocating all its elements until needed, or to expose data stored in external memory. When an ALTREP object is used in R, method hooks determine how to access its length, data pointer, or other metadata. This gives package developers a way to integrate external memory management systems while maintaining seamless integration with R semantics.

A common alternative to ALTREP is the use of file‑backed matrices, where binary files on disk are memory‑mapped into R \citep{kane2013scalable}. The OS handles paging parts of the file into memory on demand. Packages such as bigmemory or bigstatsr provide functions like big.matrix, which create a matrix stored on disk but accessible through R as if it were in memory \citep{prive2018efficient}. For example, bigstatsr allows users to run genome‑wide association studies on datasets larger than RAM by memory‑mapping genotypes from disk and performing analysis column‑wise with minimal memory footprint. File‑backed matrices offer persistence and the ability to handle datasets that exceed physical RAM, but they rely on disk I/O performance and are not as fast as in‑RAM shared memory, thus a performance sacrifice is made in cases where a single copy of the data would actually fit in RAM but not one copy for each worker. Therefore, this work focuses on ALTREP based techniques in the case of data fitting into RAM once but not multiple times. Our contributions are

\begin{enumerate}
\def\labelenumi{\arabic{enumi}.}
\item
  A fully independent and user-friendly implementation based on the ALTREP framework.
\item
  A comparison of memshare vs SharedObject: data model, safety, copy‑on‑write, and developer surface showing two times faster runtime for memshare on parallel column operations without extra RSS.
\item
  A practical template for MI‑PDE feature selection on RNA‑seq.
\end{enumerate}

\section{Background}\label{background}

\subsection{ALTREP and shared memory in R}\label{altrep-and-shared-memory-in-r}

In R, ALTREP (short for ALTernate REPresentations) is a framework introduced in version 3.5.0 that allows vectors and other objects to be stored and accessed in non-standard ways while maintaining their usual R interface. Built-in type checks cannot tell the difference between an ALTREP object and its ordinary counterpart, which ensures compatibility.

Instead of relying solely on R's default contiguous in-memory arrays, ALTREP permits objects such as integers, doubles, or strings to be backed by alternative storage mechanisms. Developers can override fundamental methods that govern vector behavior---such as length queries, element access (DATAPTR, DATAPTR\_OR\_NULL, etc.), duplication, coercion, and even printing---so that objects can behave normally while drawing data from different sources.

Because these overrides are transparent to R's higher-level functions, ALTREP objects can be passed, transformed, and manipulated like regular vectors, regardless of whether their contents reside in memory, on disk, or are computed lazily.

For package authors, this framework makes it possible to expose objects that look identical to standard R vectors but internally retrieve their data from sources like memory-mapped files, shared memory, compressed formats, or custom C++ buffers. In practice, this enables efficient handling of large datasets and unconventional data representations while keeping existing R code unchanged.

\subsection{SharedObject baseline}\label{sharedobject-baseline}

SharedObject allocates shared segments and wraps them as ALTREP \citep{sharedobject2025}. It exposes properties like copyOnWrite, sharedSubset, and sharedCopy; it supports atomic types and (with caveats) character vectors. Developers can map or unmap shared regions and query whether an object is shared. SharedObject was among the first implementations that showed how ALTREP can enable multicore parallelism by avoiding data duplication.
SharedObject provides share() to wrap an R object as a shared ALTREP, with tunables:

\begin{itemize}
\item
  copyOnWrite (default TRUE): duplicates on write; setting FALSE enables in‑place edits but is not fully supported and can lead to surprising behavior (e.g., unary minus mutating the source).
\item
  sharedSubset: whether slices are also shared; can incur extra duplication in some IDEs; often left FALSE.
\item
  sharedCopy: whether dup of a shared object remains shared.
  It supports raw, logical, integer, double, or complex and, with restrictions, character (recommended read‑only; cannot assign new, previously unseen strings). Developers can also directly allocate, map, unmap and free shared regions and query is.shared or is.altrep.
\end{itemize}

\subsection{R's threading model}\label{rs-threading-model}

R's C API is single‑threaded; package code must not call R from secondary threads. Process‑level parallelism (clusters) remains the primary avenue. Consequently, shared‑memory frameworks must ensure that mutation is either controlled in the main thread or performed at the raw buffer level without touching R internals.

\subsection{PDE-based Mutual Information}\label{pde-based-mutual-information}

For feature selection with a discrete response \(Y\) and a continuous feature \(X\), mutual information can be expressed as:
\[I(X;Y) = \sum_{y} p(y) KL(p(x|y) || p(x)),\]
requiring only univariate densities \(p(x)\) and \(p(x|y)\) per class.
This lends itself to Pareto Density Estimation (PDE), a density estimator based on hyperspheres with the Pareto radius chosen by an information-optimal criterion. In PDE, the idea is to select a subset S of the data with relative size \(p = |S|/|D|\). The information content is \(I(p) = -p*ln(p)\)). \citep{ultsch2005pareto} showed that the optimal set size corresponds to about 20.1\%, retrieving roughly 88\% of the maximum possible information. The unrealized potential (URP) quantifies deviation from the optimal set size and is minimized when \(p\approx 20\%\). For univariate density estimation this yields the Pareto radius R, which can be approximated by the 18\(\%\) quantile distance in one dimension. PDE thus adapts the neighborhood size following the Pareto rule (80--20 rule) to maximize information content. Empirical studies report that PDE can outperform standard density estimators under default settings \citep{thrun2020analyzing}.

With respect to the categorial variable no density estimation is needed as the most accurate density estimate in this case is simply the relative label count, \(p(y) = \frac{\#\{\omega\in\Omega ~|~ Y(\omega) = y\}}{\#\Omega}\). Here \(\Omega\) is the set of cases, \(Y\) is the categorial random variable and \(y\) runs over the range of \(Y\), \(Y(\Omega)\).

\section{Methods}\label{methods}

In idiomatic use, memshare coordinates PSOCK clusters (separate processes) that attach to one shared segment. Within workers the relevant shared segments are retrieved and the task is executed on it. The key win is replacing per‑worker duplication of the large matrix with a way cheaper retrieval of a handle to the shared memory segment and subsequent wrapping in an ALTREP instead.

\subsection{\texorpdfstring{The \texttt{memshare} API}{The memshare API}}\label{the-memshare-api}

Memshares source code can be found under \cite{memshareGithub}. Shared memory pages in \texttt{memshare} are handled by unique string identifiers on the OS side. These identifiers can be requested and retrieved via C/C++. To prevent two master R sessions from accidentally accessing each other's memory space because duplicate allocations can lead to undefined behavior at the OS level, users may define one or more than one \textbf{namespace} in which the current session operates.
The \texttt{memshare} API closely mirrors C's memory ownership model but applies it to R sessions. A master (primary) session owns the memory, while worker (secondary) sessions can access it.

\subsection{Shared memory semantics}\label{shared-memory-semantics}

A crucial aspect of memshare's design is how shared memory is managed and exposed through R. Three definitions clarify the terminology:

A \textbf{namespace} refers to a character string that defines the identifier of the shared memory context. It allows the initialization, retrieval, and release of shared variables under a common label, ensuring that multiple sessions or clusters can coordinate access to the same objects. While this does not provide absolute protection, it makes it the user's responsibility to avoid assigning the same namespace to multiple master sessions.

\textbf{Pages} are variables owned by the current compilation unit of the code such as the R session or terminal that loaded the DLL. Pages are realized as shared memory objects: on Windows via MapViewOfFile, and on Unix systems via shm in combination with mmap.

\textbf{Views} are references to variables owned by another or the same compilation unit. Views are always ALTREP wrappers, providing pointers to the shared memory chunk so that R can interact with them as if they were ordinary vectors or matrices.

Together, these concepts enforce a lifecycle: pages represent ownership of memory segments, views represent references to them, and namespaces serve as the coordination mechanism. The combination guarantees both memory efficiency and safety when performing multicore computations across R processes.

If the user detaches the memshare package, all handles are destroyed. This means that all variables of all namespaces are cleared, provided there is no other R thread still using them. In other words, unloading the package cleans up shared memory regions and ensures that no dangling references remain. Other threads still holding a handle to the memory will prevent this from cleaning up the memory as it would invalidate working memory of the other thread. The shared memory then is cleared whenever all handles are released.

\subsubsection{Master session}\label{master-session}

A master session takes ownership of a memory page using:

\begin{itemize}
\tightlist
\item
  \texttt{registerVariables(namespace,\ variableList)}
\end{itemize}

where \texttt{variableList} is a named list of supported types. They are double matrices, double vectors, or lists of these. The names define the memory pages ID through which they can be accessed, while the values of the actual variables define the size and content of the memory page.

To de-allocate memory pages, the master session can call:

\begin{itemize}
\tightlist
\item
  \texttt{releaseVariables(namespace,\ variableNames)}
\end{itemize}

where \texttt{variableNames} is a character vector containing the names of previously shared variables.

Memshare also allows for releasing all the memory allocated for a given namespace by a memory context, i.e., a parallel cluster with a master session, via the

\begin{itemize}
\tightlist
\item
  \texttt{memshare\_gc(namespace,\ cluster)}
\end{itemize}

function. This removes every view handle in the context first and afterwards calls to release all pages.

\textbf{Note.} Memory pages are not permanently deallocated if another session still holds a view of them. This ensures stability: allowing workers to continue with valid but outdated memory is safer than letting them access invalidated memory. However, releasing variables still in use is always a user error and must be avoided.

\subsubsection{Worker session}\label{worker-session}

Once the master session has shared variables, worker sessions can retrieve them via:

\begin{itemize}
\tightlist
\item
  \texttt{retrieveViews(namespace,\ variableNames)}
\end{itemize}

This returns a named list of R objects. These objects are raw ALTREP objects indistinguishable from the originals (\texttt{is.matrix}, \texttt{is.numeric}, etc. all behave the same).

When operating on these objects, workers interact directly with the underlying C buffer, backed by \texttt{mmap} (Unix) or \texttt{MapViewOfFile} (Windows). Changes to such objects modify the shared memory for all sessions. In this framework, however, modification is secondary---the main goal is to transfer data from the master to worker sessions.

For metadata access without retrieving views, workers can call:

\begin{itemize}
\tightlist
\item
  \texttt{retrieveMetadata(namespace,\ variableName)}
\end{itemize}

which provides information for a single variable.

After processing, workers must return their views to allow memory release by calling:

\begin{itemize}
\tightlist
\item
  \texttt{releaseViews(namespace,\ variableNames)}
\end{itemize}

\begin{figure}
\includegraphics[width=1\linewidth]{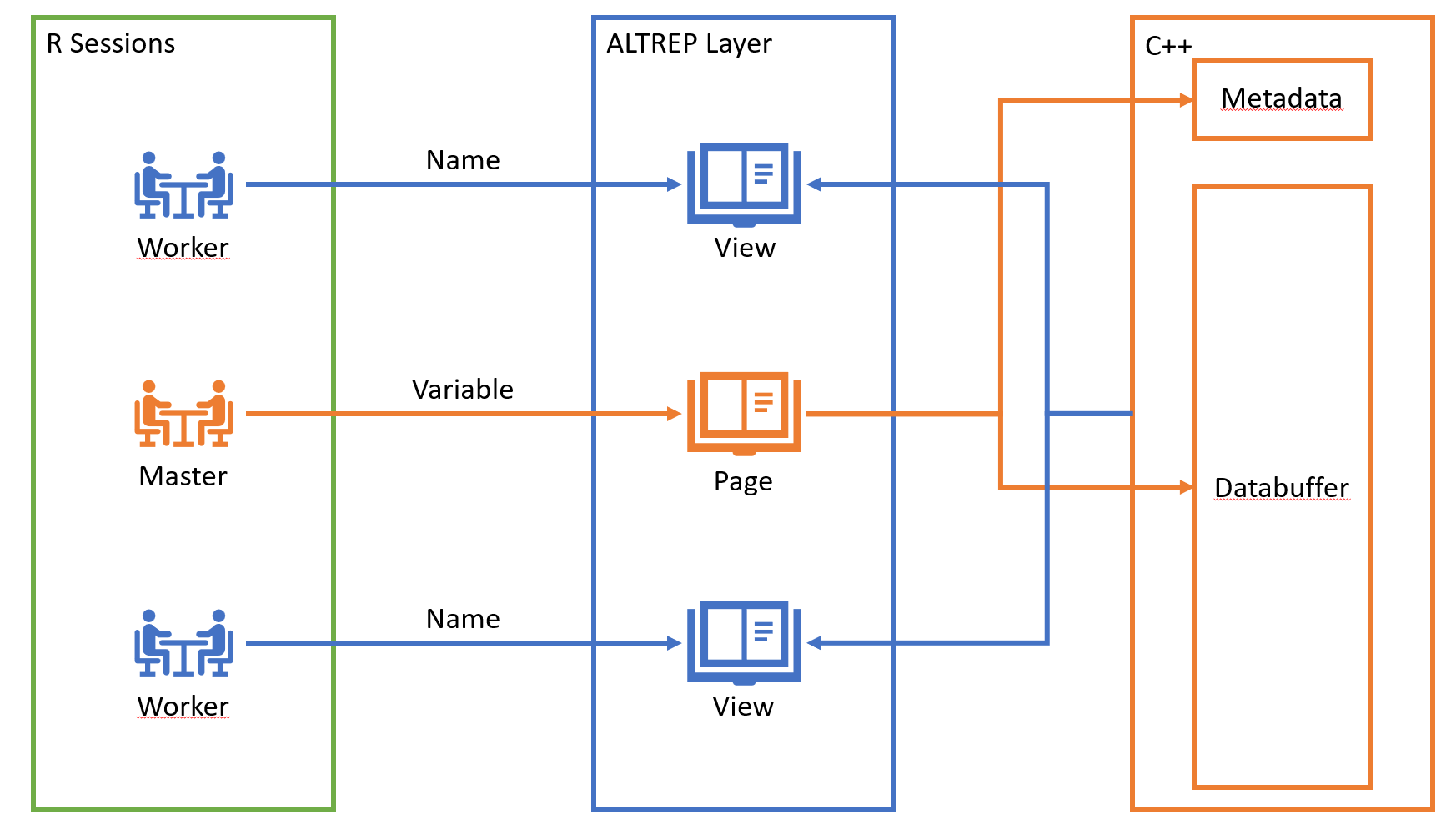} \caption{A schematic about where the memory is located and how different sessions access it.}\label{fig:figurememshare}
\end{figure}

The overall high-level concept is summarized in figure \ref{fig:figurememshare}.

\subsubsection{Diagnostic tools}\label{diagnostic-tools}

To verify correct memory management, two diagnostic functions are available:

\begin{itemize}
\tightlist
\item
  \texttt{pageList()} : lists all variables owned by the current session.\\
\item
  \texttt{viewList()} : lists all views (handles) currently held by the session.
\end{itemize}

The former is stricter, since it identifies ownership, whereas the latter only tracks held views.

\subsubsection{\texorpdfstring{User-friendly Wrapper functions for \texttt{apply} and \texttt{lapply}}{User-friendly Wrapper functions for apply and lapply}}\label{user-friendly-wrapper-functions-for-apply-and-lapply}

Since memory ownership and address sharing are low-level concepts, \texttt{memshare} provides wrapper functions that mimic \texttt{parallel::parApply} and \texttt{parallel::parLapply}.

\begin{itemize}
\tightlist
\item
  \texttt{memApply(X,\ MARGIN,\ FUN,\ NAMESPACE,\ CLUSTER,\ VARS,\ MAX.CORES)}\strut \\
  Mimics \texttt{parallel::parApply}.

  \begin{itemize}
  \tightlist
  \item
    \texttt{X}: a double matrix.\\
  \item
    \texttt{MARGIN}: direction (\texttt{1\ =\ row-wise}, \texttt{2\ =\ column-wise}).\\
  \item
    \texttt{FUN}: function applied to rows/columns.\\
  \item
    \texttt{CLUSTER}: a prepared parallel cluster (variables exported via \texttt{parallel::clusterExport}).\\
  \item
    \texttt{VARS}: additional shared variables (names must match \texttt{FUN} arguments).\\
  \item
    \texttt{MAX.CORES}: only relevant if \texttt{CLUSTER} is uninitialized.
  \end{itemize}
\end{itemize}

\texttt{memApply} automatically manages sharing and cleanup of \texttt{X} and \texttt{VARS}, ensuring no residual C/C++ buffers remain. Both \texttt{X} and \texttt{VARS} can also refer to previously allocated shared variables, though in that case the user must manage their lifetime.

\begin{itemize}
\tightlist
\item
  \texttt{memLapply(X,\ FUN,\ NAMESPACE,\ CLUSTER,\ VARS,\ MAX.CORES)}\strut \\
  Equivalent to \texttt{parallel::parLapply}, but within the \texttt{memshare} framework.

  \begin{itemize}
  \tightlist
  \item
    \texttt{X}: a list of double matrices or vectors.\\
  \item
    Other arguments behave the same way as in \texttt{memApply}.
  \end{itemize}
\end{itemize}

\subsection{Examples of Use}\label{examples-of-use}

We provide two top-level examples for the use of the memshare package. One with memLapply and one with memApply.

The first example computes the correlation between each column of a matrix and a reference vector using shared memory and memApply. The matrix can be provided directly (and will be registered automatically) or by name if already registered.

\begin{verbatim}
library(memshare)
set.seed(1)
n <- 10000
p <- 2000
# Numeric double matrix (required): n rows (cases) x d columns (features)
X <- matrix(rnorm(n * p), n, p)
# Reference vector to correlate with each column
y <- rnorm(n)
f = function(v, y) cor(v, y)

ns <- "my_namespace"
res <- memshare::memApply(
X = X, MARGIN = 2,
FUN = f,
NAMESPACE = ns,
VARS = list(y = y),
MAX.CORES = NULL # defaults to detectCores() - 1
)
\end{verbatim}

memApply parallelizes a row- or column-wise map over a matrix that lives once in shared memory. If X is passed as an ordinary R matrix, it is registered under a generated name in the namespace ns. Additional variables (here y) can be provided as a named list; these are registered and retrieved as ALTREP views on the workers. A cluster is created automatically if none is provided. Each worker obtains a cached view of the matrix (and any shared variables), extracts the i-th row or column as a vector v according to MARGIN, calls FUN(v,\ldots), and returns the result. Views are released after the computation, and any objects registered by this call are freed. Because workers operate on shared views rather than copies, the total resident memory remains close to a single in-RAM copy of X, while runtime scales with the available cores.

As a second example, consider a case where a list of 1000 random matrices is multiplied by a random vector. This task is parallelizable at the element level and demonstrates the use of memshare::memLapply, which applies a function across list elements in a shared memory context:

\begin{verbatim}
  library(memshare)
  list_length = 1000
  matrix_dim = 100
  
  # Create the list of random matrices
  l <- lapply(
      1:list_length,
      function(i) matrix(rnorm(matrix_dim * matrix_dim),
      nrow = matrix_dim, ncol = matrix_dim))
  
  y = rnorm(matrix_dim)
  
  namespace = "my_namespace"
  res = memLapply(l, function(el, y) {
    el %*% y
  }, NAMESPACE=namespace, VARS=list(y=y), MAX.CORES = 1)#MAX.CORES=1 for simplicity
\end{verbatim}

memLapply() provides a parallel version of lapply() where the list elements and optional auxiliary variables are stored in shared memory. If the input X is an ordinary R list, it is first registered in a shared memory namespace. Additional variables can be supplied either as names of existing shared objects or as a named list to be registered. A parallel cluster is created automatically if none is provided, and each worker is initialized with the memshare environment.

For each index of the list, the worker retrieves an ALTREP view of the corresponding element (and of any shared variables), applies the user-defined function FUN to these objects, and then releases the views to avoid memory leaks. The function enforces that the first argument of FUN corresponds to the list element and that the names of shared variables match exactly between the namespace and the function signature. Results are collected with parLapply, yielding an ordinary R list of the same length as the input.

Because only lightweight references to the shared objects are passed to the workers, no duplication of large data occurs, making the approach memory-efficient. Finally, memLapply() includes cleanup routines to release temporary registrations, stop the cluster if it was created internally, and free shared memory, ensuring safe reuse in subsequent computations.

\subsection{Benchmark design}\label{benchmark-design}

We compare memshare and SharedObject on a column‑wise apply task computing per column across square matrices of sizes for \(10^i x 10^i\) i in 1,\ldots,5. We use a PSOCK cluster with detectCores()-1 on an iMac PRO, 256 GB DDR4, 2,3 GHz 18-Core Intel Xeon W. For each size, we run 100 repetitions and record wall‑clock times and resident set size (RSS) across all worker PIDs plus the master. The RSS is summed via ps and our helper total\_rss\_mb(). For SharedObject we create A2 \textless- share(A1)) and parApply(); for memshare we call memApply directly on A1 with a namespace, so that only ALTREP views are created on workers. A serial baseline uses apply. A minimally edited version of the full script (setup, PID collection, loops, and data saving) is provided in Appendix A to ensure reproducibility.

As part of our safety and lifecycle checks, we ensure that views, which keep shared segments alive, are always released in the workers before returning control. Once all work is complete, the corresponding variables are then released in the master. To maintain fairness, we avoid creating incidental copies, such as those introduced by coercions, remove variables with rm() and use R's garbage collection gc() after each call.

\subsection{RNA-seq dataset via FireBrowse}\label{rna-seq-dataset-via-firebrowse}

FireBrowse, \citep{firebrowse2025}, delivers raw counts of gene expression indexed by NCBI identifiers. For each gene identifier \(i\) (from Ensembl or NCBI), we obtain a raw count \(r_i\) that quantifies the observed read abundance. These raw counts represent the number of reads mapped to each gene, without length normalization. To convert raw counts into TPM (Transcripts Per Million, \citep{li2010rsem}), we require gene or transcript lengths \(l_i\). For each gene \(i\), we compute:

\[\hat{r}_i=\frac{r_i}{l_i}\]
The total sum \(R = \sum_i \hat{r}_i\) across all genes is then used to scale values as:
\[TPM_i=\frac{\hat{r}_i}{R} \times 10^6\]
This transformation allows comparison of expression levels across genes and samples by correcting for gene length and sequencing depth \citep{li2010rsem}. After transformation, our dataset consists of \(d = 19,637\) gene expressions across \(N = 10,446\) cases spanning 32 diagnoses. It can be found under \citep{thrun2025genexpressions}.

\section{Results}\label{results}

In the first subsection the efficiency of memshare is compared to ShareObject and in the second subsection the application is presented.

\subsection{Performance \& memory}\label{performance-memory}

\begin{figure}
\includegraphics[width=1\linewidth,height=0.3\textheight]{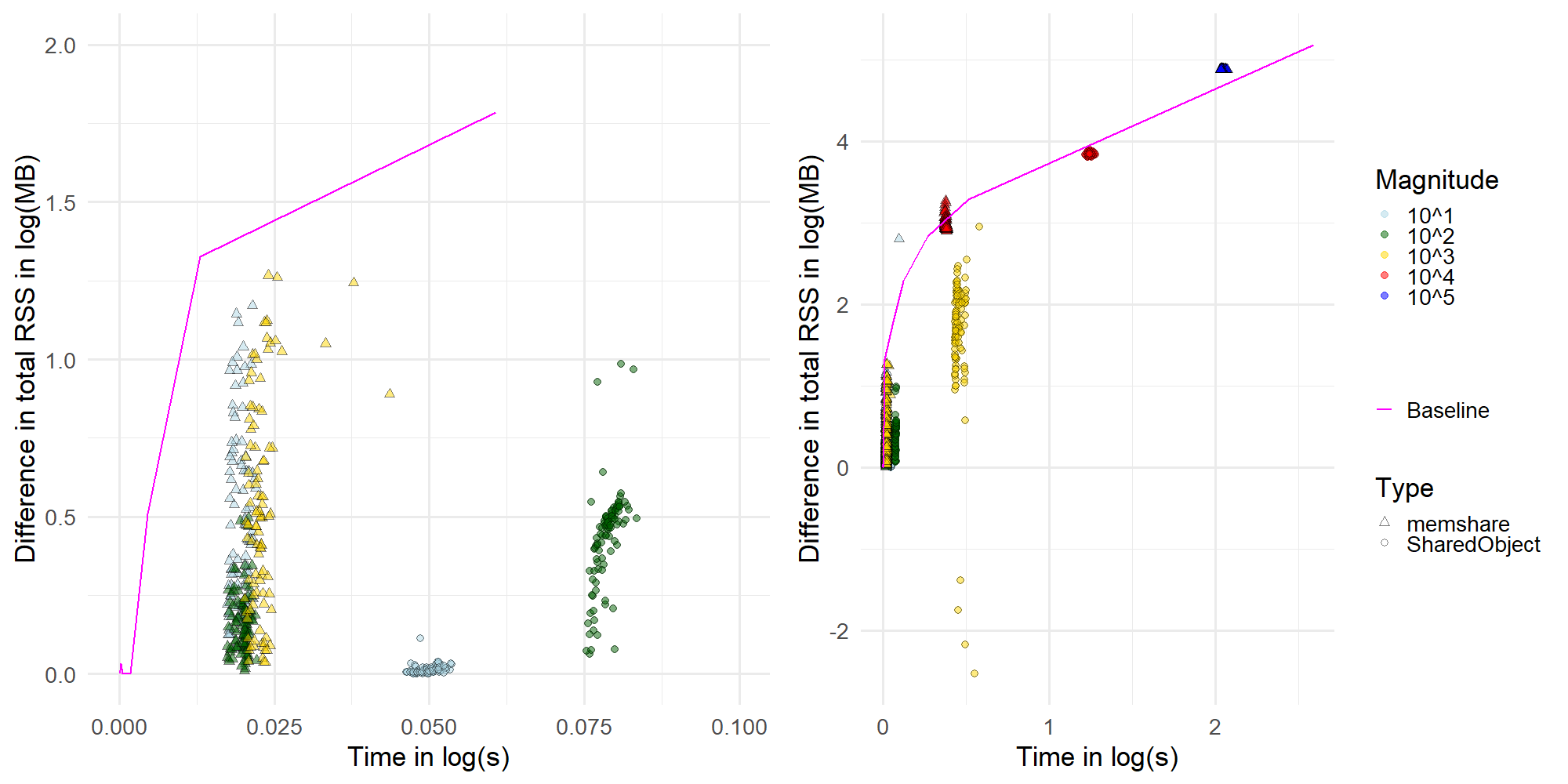} \caption{Median runtime (log‑scale) vs matrix size for memshare, SharedObject, and serial baseline; ribbons show IQR across 100 runs. Insets show total RSS (MB) during the run relative to idle.}\label{fig:figure1}
\end{figure}

In figure \ref{fig:figure1}, the results are presented for square matrices of increasing magnitudes. The left panel shows detailed scatter plots for the first three matrix sizes, while the right panel summarizes all five magnitudes from \(10^1\) to \(10^5\). The x-axis represents computation time (log seconds), and the y-axis represents memory consumption (log megabytes). It is measured as total RSS. Each point corresponds to one of 100 trials per magnitude. The magenta baseline indicates the performance of a single-threaded R computation.

Memshare (triangles) consistently outperforms SharedObject (circles). For the first three magnitudes, both packages exhibit lower memory consumption than the baseline. At \(10^4\), average memory usage is comparable to the baseline, while at \(10^5\), Memshare slightly exceeds it. SharedObject, however, could not be executed for this magnitude (see Appendix B).

Table \ref{tab:median-res-tab-static} reports the median values, and the variance is estimated using the amad statistic (Table \ref{tab:amad-res-tab-static}) of figure \ref{fig:figure1}. Considering relative differences \citep{Ultsch2008}, Memshare achieves computation times that are 90--170\% faster than SharedObject. For matrices of size \(10^2\) and larger, memory consumption is reduced by 132--153\% compared to SharedObject.

\begin{table}
\centering
\caption{\label{tab:median-res-tab-static}The benchmark compares three types: memshare, SharedObject, and baseline. For memshare and SharedObject, the reported values are the medians over 100 iterations, while the baseline result comes from a single-threaded R run using one iteration. Exponent refers to the matrix size, sec to time difference measurement in seconds, and the forth column to difference in memory size in megabyte.}
\centering
\fontsize{7}{9}\selectfont
\begin{tabular}[t]{l|r|r|r|r}
\hline
Type & Exponent & Diff.Sec & MemoryDiff.MB & mem.after.call\\
\hline
SharedObject & 1 & 0.121 & 0.021 & 3226.453\\
\hline
SharedObject & 2 & 0.199 & 1.961 & 3441.057\\
\hline
SharedObject & 3 & 1.812 & 56.824 & 10742.508\\
\hline
SharedObject & 4 & 16.614 & 6909.932 & 41330.389\\
\hline
SharedObject & 5 & NaN & NaN & NaN\\
\hline
memshare & 1 & 0.046 & 1.676 & 3612.117\\
\hline
memshare & 2 & 0.048 & 0.400 & 3709.770\\
\hline
memshare & 3 & 0.053 & 2.139 & 4005.373\\
\hline
memshare & 4 & 1.386 & 927.875 & 22821.143\\
\hline
memshare & 5 & 108.151 & 76312.072 & 114774.424\\
\hline
Baseline & 1 & 0.000 & 0.023 & 2379.555\\
\hline
Baseline & 2 & 0.001 & 0.000 & 2379.672\\
\hline
Baseline & 3 & 0.030 & 20.270 & 2409.809\\
\hline
Baseline & 4 & 2.275 & 1949.070 & 6369.062\\
\hline
Baseline & 5 & 386.979 & 152752.324 & 232890.570\\
\hline
\end{tabular}
\end{table}

\begin{table}
\centering
\caption{\label{tab:amad-res-tab-static}AMAD for the benchmark of SharedObject vs memshare.}
\centering
\fontsize{7}{9}\selectfont
\begin{tabular}[t]{l|l|r|r|r|r}
\hline
  & Type & Exponent & Diff.Sec & MemoryDiff.MB & mem.after.call\\
\hline
1 & SharedObject & 1 & 0.008 & 0.019 & 1.043\\
\hline
2 & SharedObject & 2 & 0.008 & 0.779 & 94.434\\
\hline
3 & SharedObject & 3 & 0.130 & 82.204 & 1115.637\\
\hline
4 & SharedObject & 4 & 0.524 & 282.854 & 8959.919\\
\hline
10 & SharedObject & 5 & NaN & NaN & NaN\\
\hline
5 & memshare & 1 & 0.005 & 2.225 & 93.629\\
\hline
6 & memshare & 2 & 0.003 & 0.361 & 21.258\\
\hline
7 & memshare & 3 & 0.005 & 3.106 & 104.097\\
\hline
8 & memshare & 4 & 0.052 & 144.659 & 7173.754\\
\hline
9 & memshare & 5 & 2.522 & 179.687 & 1944.821\\
\hline
\end{tabular}
\end{table}

\subsection{Application to Feature Selection by Mutual Information using Pareto Density Estimation}\label{application-to-feature-selection-by-mutual-information-using-pareto-density-estimation}

\begin{figure}
\includegraphics[width=1\linewidth]{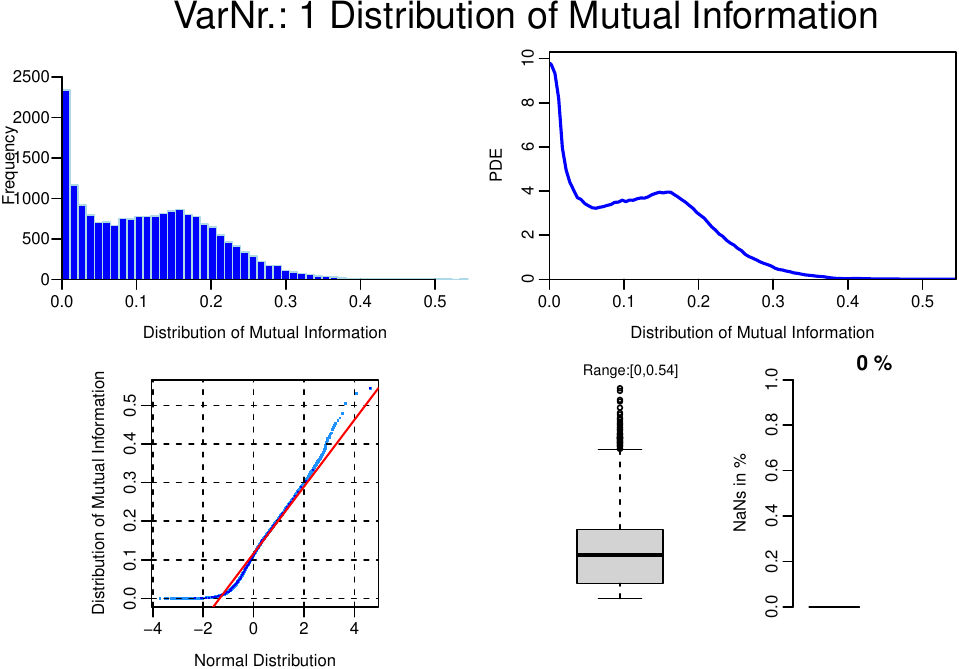} \caption{The distribution of mutual information for 19637 gene expressions as a histogram, pareto density estimation (PDE), QQ-plot against normal distribution and boxplot. There are not missing values (NaN).}\label{fig:figure2}
\end{figure}

The computation of mutual information produced values ranging from 0 to 0.54 (Figure \ref{fig:figure2}). The QQ-plot shows clear deviation from a straight line, indicating that the distribution is not Gaussian. Both the histogram and the PDE plot provide consistent estimates of the probability density, revealing a bimodal structure. The boxplot further highlights the presence of outliers with values above 0.4.

The analysis required about two hours of computation time and approximately 47 GB of RAM---feasible only through memory sharing. In practice, mutual information values can guide feature selection, either by applying a hard threshold or by using a soft approach via a mixture model, depending on the requirements of the subsequent machine learning task.

\section{Discussion}\label{discussion}

Across matrix sizes, Memshare achieved a two-fold reduction in median computation time compared to SharedObject on the column-wise task. For matrices of size \(10^2\) and larger, memory consumption measured by total RSS was between half and a third of that of ShareObject. At the smallest size (\(10^1\)), Memshare consumed more memory than SharedObject because, in R, the metadata must also be shared; this requires a second shared-memory segment whose fixed overhead dominates at small sizes.

The RS of all worker processes plus the master process remained in Memshare approximately the same as for the baseline, indicating the absence of redundant copies. In contrast, SharedObject exhibited overhead consistent with copy-on-write materializations and temporary object creation up to size \(10^4\). Its memory usage was approximately an order of magnitude higher than that of Memshare or the baseline, as illustrated by the triangles aligning with the baseline of a higher magnitude in Figure \ref{fig:figure1}. For matrices of size \(10^5\), SharedObject caused RStudio to crash (see appendix B), suggesting a memory leak at this scale.

Overall, Memshare provides a more stable memory-sharing interface, scales more effectively to large matrix sizes, and achieves greater computational efficiency than SharedObject.

\section{Summary}\label{summary}

Regardless of package, R's single‑threaded API implies that multi‑threaded computation should touch only raw memory and synchronize results at the main thread boundary. Shared mutation requires external synchronization if multiple workers write to overlapping regions. In practice, read‑mostly pattern are ideal.

Here, memshare's namespace + view model and memApply wrapper simplify cross‑process sharing compared to manual share()) + cluster wiring. Its explicit releaseViews/Variables lifecycle makes retention and cleanup auditable. SharedObject's fine‑grained properties are powerful, but the interaction of copy‑on‑write and duplication semantics increases cognitive load.

memshare combines ALTREP‑backed shared memory with a pragmatic parallel API to deliver strong speed and memory efficiency on multicore systems. In analytic pipelines like MI‑based feature selection for RNA‑seq, this enables simple, scalable patterns---one in‑RAM copy of the data, many cores, and no serialization overhead.

\section{Appendix A: code listing of benchmark}\label{appendix-a-code-listing-of-benchmark}

To avoid the crash message for ShareObject in appendix B, we tried manual garbage collection and performed saves for each iteration without beeing able to change the outcome.

\begin{verbatim}
#01BenchmarkMemshareVSSharedObject.R
comment="01BenchmarkMemshareVSSharedObject.R"

library(ps)

# Get the PIDs of the workers in a PSOCK cluster
cluster_pids = function(cl) {
  as.integer(unlist(clusterCall(cl, Sys.getpid)))
}

# Sum RSS (MB) for a vector of PIDs; optionally include the master process
total_rss_mb = function(pids, include_master = TRUE) {
  rss_pid = function(pid) {
    h = ps_handle(pid)
    as.numeric(ps_memory_info(h)["rss"]) / (1024^2)
  }
  # sum worker RSS; skip PIDs that may have exited
  worker_sum = sum(vapply(pids, function(pid) {
    tryCatch(rss_pid(pid), error = function(e) 0)
  }, numeric(1)))
  
  if (include_master) {
    master = as.numeric(ps_memory_info()["rss"]) / (1024^2)
    worker_sum + master
  } else {
    worker_sum
  }
}

library(parallel)
cl=makeCluster(detectCores()-1)
pids = cluster_pids(cl)
set.seed(1234)
Ivec=seq(from=1,to=5,by=1)

##SharedObject ----
library(SharedObject)
mem_idle_SharedObject = total_rss_mb(pids, include_master = TRUE)

SharedObjectPerformance=c()

for(i in Ivec){
  if(i>1){
    save(file=file.path(RelPath(1,"01Transformierte"),"SharedObjectPerformance_v2.rda"),
         SharedObjectPerformance,comment,mem_idle_SharedObject)
  }
  Start=c()
  Ende=c()
  print(i)
  mem_before_call=c()
  mem_after_call=c()
  for(k in 1:100){
      y=runif(1:10^(i+1))
      mem_before_call[k] = total_rss_mb(pids, include_master = TRUE)
      x1=Sys.time()
      A1 = matrix(y,10^i, 10^i)
      A2 = share(A1)
      res=parApply(cl,X = A2,FUN = sd,MARGIN = 2)
      Start[k]=x1
      freeSharedMemory(listSharedObjects())
      x2=Sys.time()
      Ende[k]=x2
      mem_after_call[k] = total_rss_mb(pids, include_master = TRUE)
      rm(A1)
      rm(y)
      gc() 
      save(file=file.path(RelPath(1,"01Transformierte"),paste0("SharedObjectPerformance_v2_part",i,".rda")),
           mem_before_call,mem_after_call,Start,Ende,comment)
  }
  #temporary save ----
  mem_at_end_SharedObject = total_rss_mb(pids, include_master = TRUE)
  
  SharedObjectPerformance[[i]]=cbind(DiffTime=Ende-Start,Start,Ende,mem_before_call,mem_after_call)
  
  save(file=file.path(RelPath(1,"01Transformierte"),"SharedObjectPerformance_v3.rda"),
       SharedObjectPerformance,comment,mem_idle_SharedObject,mem_at_end_SharedObject)
  
i=i+1  
}
mem_at_end_SharedObject = total_rss_mb(pids, include_master = TRUE)

save(file=file.path(RelPath(1,"01Transformierte"),"SharedObjectPerformance_v3.rda"),
     SharedObjectPerformance,comment,mem_idle_SharedObject,mem_at_end_SharedObject)

listSharedObjects()
rstudioapi::restartSession(clean = T)

##memshare ----

library(memshare)
mem_idle_memshare = total_rss_mb(pids, include_master = TRUE)
memsharePerformance=c()
for(i in  Ivec){
  Start=c()
  Ende=c()
  mem_before_call=c()
  mem_after_call=c()
  print(i)
  for(k in 1:100){
    y=runif(1:10^(i+1))
    mem_before_call[k] = total_rss_mb(pids, include_master = TRUE)
    x1=Sys.time()
    A1 = matrix(y,10^i, 10^i)
    res = memshare::memApply(X = A1, MARGIN = 2, FUN = function(x) {
      return(sd(x))
    }, CLUSTER=cl, NAMESPACE="namespace_id")
    #memshare::releaseViews("my_namespace",A1)
    x2=Sys.time()
    Start[k]=x1
    Ende[k]=x2
    mem_after_call[k] = total_rss_mb(pids, include_master = TRUE)
    rm(A1)
    rm(y)
    gc() 
  }
  memsharePerformance[[i]]=cbind(DiffTime=Ende-Start,Start,Ende,mem_before_call,mem_after_call)
}
mem_at_end_memshare = total_rss_mb(pids, include_master = TRUE)

save(file=file.path(RelPath(1,"01Transformierte"),"memsharePerformance.rda"),
     memsharePerformance,comment,mem_idle_memshare,mem_at_end_memshare)
rstudioapi::restartSession(clean = T)
\end{verbatim}

\section{Appendix B: screenshot}\label{appendix-b-screenshot}

\begin{figure}[H]
\centering
\includegraphics[width=.6\linewidth]{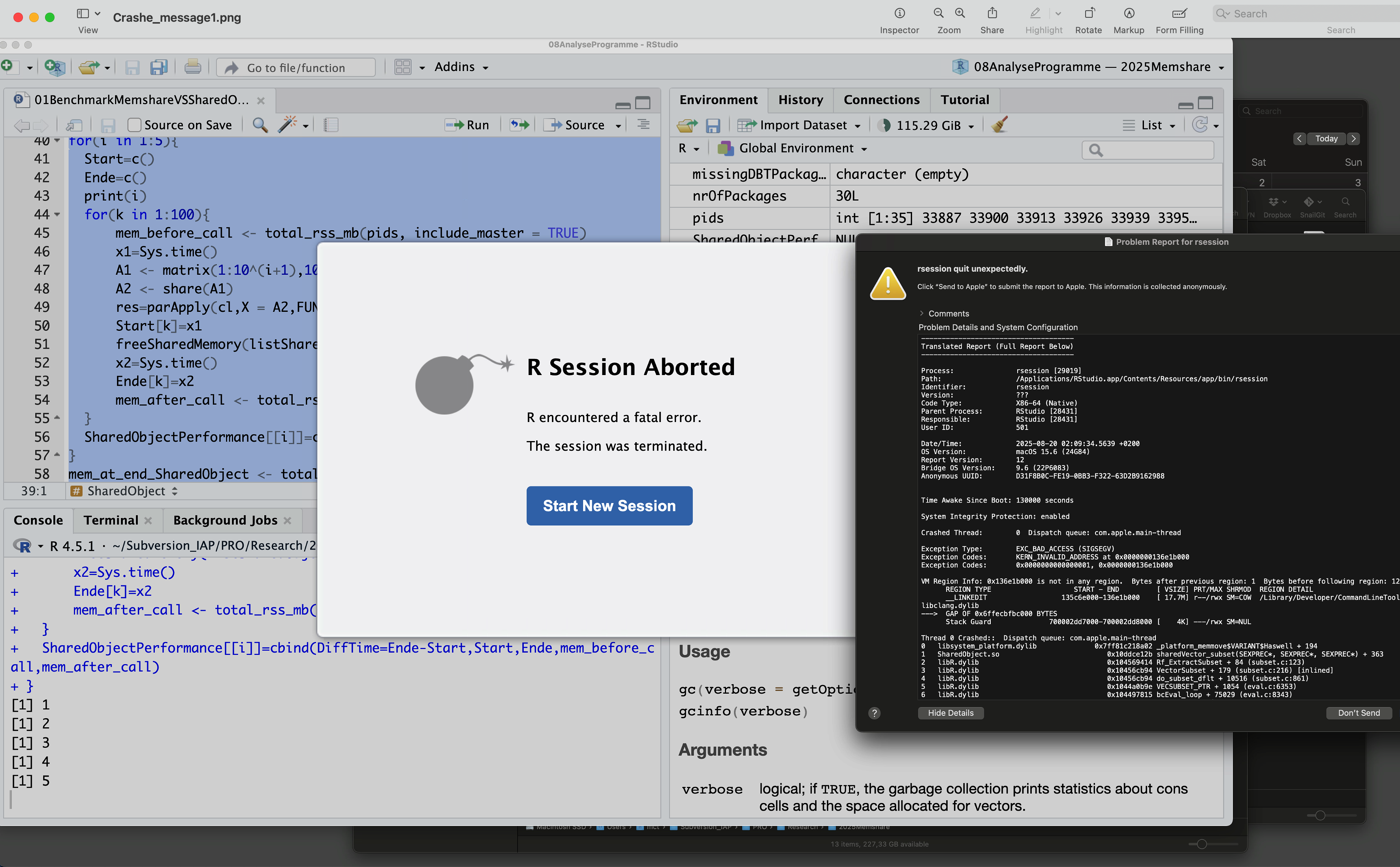} \caption{First Screenshot of ShareObjects Computation.}\label{fig:app-a-1}
\end{figure}

\begin{figure}[H]
\centering
\includegraphics[width=.6\linewidth]{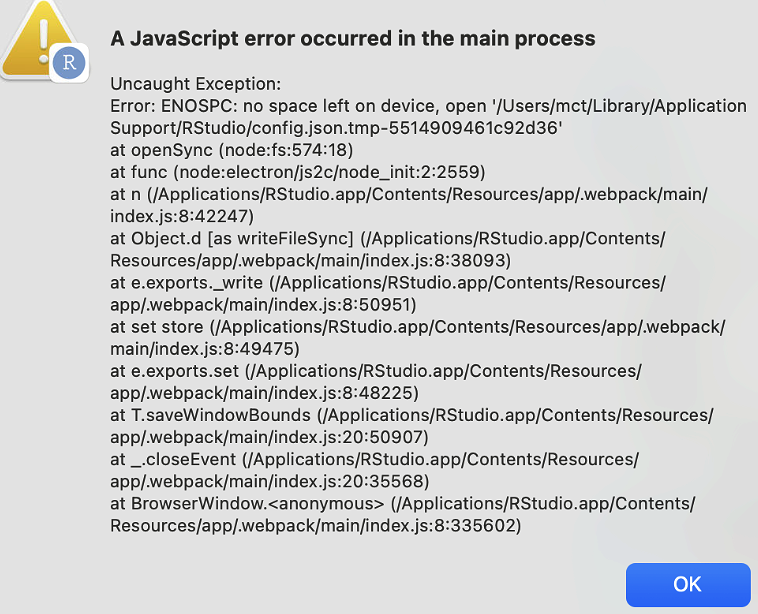} \caption{Second Screenshot of ShareObjects Computation.}\label{fig:app-a-2}
\end{figure}

Report as screenshots im figure \ref{fig:app-a-1} and subsequent after forcing to close Rstudio in \ref{fig:app-a-2} of the crash of Rstudio if ShareObject is called with matrix of size \(10^5\).

\bibliography{RJreferences}

\begin{thebibliography}{13}
\providecommand{\natexlab}[1]{#1}
\providecommand{\url}[1]{\texttt{#1}}
\expandafter\ifx\csname urlstyle\endcsname\relax
  \providecommand{\doi}[1]{doi: #1}\else
  \providecommand{\doi}{doi: \begingroup \urlstyle{rm}\Url}\fi

\bibitem[{Broad Institute of MIT and Harvard}(2025)]{firebrowse2025}
{Broad Institute of MIT and Harvard}.
\newblock Firebrowse (rrid:scr\_026320).
\newblock \url{http://firebrowse.org/}, 2025.
\newblock Accessed via \url{https://gdac.broadinstitute.org}.

\bibitem[Corporation and Weston(2025)]{doparallel2025}
Microsoft Corporation and Steve Weston.
\newblock \emph{doParallel: Foreach Parallel Adaptor for the 'parallel' Package}, 2025.
\newblock URL \url{https://github.com/revolutionanalytics/doparallel}.
\newblock R package version 1.0.17.

\bibitem[Kane et~al.(2013)Kane, Emerson, and Weston]{kane2013scalable}
Michael Kane, John~W. Emerson, and Stephen Weston.
\newblock Scalable strategies for computing with massive data.
\newblock \emph{Journal of Statistical Software}, 55\penalty0 (14):\penalty0 1--19, 2013.
\newblock \doi{10.18637/jss.v055.i14}.
\newblock URL \url{https://doi.org/10.18637/jss.v055.i14}.

\bibitem[Li and Dewey(2011)]{li2010rsem}
Bo~Li and Colin~N. Dewey.
\newblock Rsem: accurate transcript quantification from rna-seq data with or without a reference genome.
\newblock \emph{BMC Bioinformatics}, 12\penalty0 (1):\penalty0 323, 2011.
\newblock \doi{10.1186/1471-2105-12-323}.
\newblock URL \url{https://doi.org/10.1186/1471-2105-12-323}.

\bibitem[{Michael C. Thrun and Julian Märte}(2025)]{memshareGithub}
{Michael C. Thrun and Julian Märte}.
\newblock memshare.
\newblock \url{https://github.com/Mthrun/memshare}, 2025.

\bibitem[Märte and Thrun(2025)]{thrunmaerteR}
Julian Märte and Michael~C. Thrun.
\newblock \emph{memshare}, 2025.
\newblock URL \url{https://cran.r-project.org/package=memshare}.
\newblock CRAN published R package.

\bibitem[Prive et~al.(2018)Prive, Aschard, Ziyatdinov, and Blum]{prive2018efficient}
Florian Prive, Hugues Aschard, Andrey Ziyatdinov, and Michael G.~B. Blum.
\newblock Efficient analysis of large-scale genome-wide data with two r packages: bigstatsr and bigsnpr.
\newblock \emph{Bioinformatics}, 34\penalty0 (16):\penalty0 2781--2787, 2018.
\newblock \doi{10.1093/bioinformatics/bty185}.
\newblock URL \url{https://doi.org/10.1093/bioinformatics/bty185}.

\bibitem[Team(2025)]{Rparallel2025}
R~Core Team.
\newblock \emph{Support for Parallel Computation in R}, 2025.
\newblock URL \url{https://stat.ethz.ch/R-manual/R-devel/library/parallel/doc/parallel.pdf}.
\newblock R package 'parallel' version included in R.

\bibitem[Thrun and Märte(2025)]{thrun2025genexpressions}
Michael~C. Thrun and Julian Märte.
\newblock Genexpressions dataset derived from firebrowse, 2025.
\newblock URL \url{https://zenodo.org/records/16937028}.

\bibitem[Thrun et~al.(2020)Thrun, Gehlert, and Ultsch]{thrun2020analyzing}
Michael~C. Thrun, Tim Gehlert, and Alfred Ultsch.
\newblock Analyzing the fine structure of distributions.
\newblock \emph{PLOS ONE}, 15\penalty0 (10):\penalty0 e0238835, 2020.
\newblock \doi{10.1371/journal.pone.0238835}.
\newblock URL \url{https://doi.org/10.1371/journal.pone.0238835}.

\bibitem[Ultsch(2005)]{ultsch2005pareto}
Alfred Ultsch.
\newblock Pareto density estimation: A density estimation for knowledge discovery.
\newblock In \emph{Proceedings of the 28th Annual Conference of the German Classification Society (GfKl)}, Studies in Classification, Data Analysis, and Knowledge Organization, pages 91--98. Springer, 2005.

\bibitem[Ultsch(2008)]{Ultsch2008}
Alfred Ultsch.
\newblock Is log ratio a good value for measuring return in stock investments?
\newblock In \emph{Advances in Data Analysis, Data Handling and Business Intelligence}, Studies in Classification, Data Analysis, and Knowledge Organization, pages 505--511. Springer, 2008.

\bibitem[Wang and Morgan(2025)]{sharedobject2025}
Jiefei Wang and Martin Morgan.
\newblock \emph{SharedObject: Sharing R objects across multiple R processes without memory duplication}, 2025.
\newblock URL \url{https://bioconductor.org/packages/SharedObject}.
\newblock R package version (Bioconductor Release).

\end{thebibliography}

\end{document}